\documentclass[letterpaper]{aa}
\usepackage{graphics}
\usepackage{txfonts}
\usepackage{natbib}
\def\msun{{\rm\,M_\odot}}
\def\be{\begin{equation}}
\def\ee{\end{equation}}

\begin{document}

\title
{The Cosmological History of Accretion onto Dark Halos and Supermassive Black Holes}
\subtitle{}

\titlerunning{Cosmological accretion history}

\author
{L.\ Miller\inst{1} \and W.\ J.\ Percival\inst{2,3} \and S.\ M.\ Croom\inst{4}
\and A. Babi{\'c}\inst{1}}

\authorrunning{L.\ Miller et al.\ }

\institute{Dept. of Physics, University of Oxford, 
Denys Wilkinson Building, Keble Road, Oxford OX1 3RH, U.K.
\and Institute for Astronomy,
University of Edinburgh, Blackford Hill, Edinburgh EH9 3HJ, U.K.
\and
Institute of Cosmology and Gravitation,
University of Portsmouth,
Portsmouth, PO1 2EG, U.K.
\and 
Anglo-Australian Observatory, PO Box 296, Epping, NSW 2121, Australia.
}

\date{Received / Accepted}

\abstract{}
{
We investigate the cosmological growth of dark halos and follow
the consequences of coeval growth for the accretion
history of associated supermassive black holes.}
{The Press-Schechter approximation is used to obtain an
analytic expression for the mean rate of growth of dark matter halos.  
Dark halo accretion
rates are compared with numerical work and the consequences
for understanding AGN evolution are described.
}  
{The mean accretion rate onto dark matter halos is shown to have a
simple analytic form that agrees with previous numerical work and that
may easily be calculated for a wide range of halo mass, redshift and
cosmological parameters.  The result offers a significant improvement over
published fitting formulae deduced from merger trees.
We then consider the growth of associated 
supermassive black holes, and make a basic test of  
the simple hypothesis of ``Pure Coeval Evolution'' (PCE) in which,
{\em on average}, black
hole growth tracks dark halo growth.  We demonstrate that 
both the absolute value of the integrated AGN bolometric luminosity density
and its cosmological evolution derived from hard X-ray surveys
are well-reproduced by PCE.
Excellent agreement is found at $z \ga 0.5$, although the observed luminosity
density drops by a factor 2 compared with PCE by $z=0$: 
black hole growth appears to decouple from halo growth at low redshifts,
and this may be related to the phenomenon of ``cosmic downsizing''.
Overall, AGN evolution appears either to be caused by or to be closely linked to 
the slow-down in the growth of cosmic structure. 
We also discuss the mean Eddington ratio averaged over all 
galaxies, which is predicted to show strong evolution to higher values
with redshift.
}
{}


\maketitle

\section{Introduction}

One of the great mysteries in the study of active galaxies and QSOs is
the physical origin of the strong cosmological evolution in their
space density at a given luminosity.  For many years it has appeared
that, at least in broad terms, the evolution was best described as
``pure luminosity evolution'' in which QSOs appeared on average to
have dimmed with time rather than changing in space density
\citep{marshall,boyle88,croom04}.  This is in accord with recent
evidence that almost all galaxies with a massive spheroid component
contain a supermassive black hole at the present epoch
\citep{magorrian,ferrarese00,gebhardt00,tremaine,onken}, 
the great majority of which must
be largely inactive today.  If these black holes were luminous in a
phase of accretion and growth at higher redshift then their
luminosities must indeed have declined with time.

Yet this picture is not easily in accord with modern ideas of galaxy formation,
since if the matter content of the universe is dominated by cold dark matter
we expect galaxies to grow hierarchically, and we expect the mass function of
black holes at the centres of galaxies to increase with time - what then is
the mechanism that allows the mass function to increase but the luminosity to
decrease with cosmic epoch?  One obvious possibility is that black holes may
increase in mass but with a decreasing mean accretion rate.  This picture has
been incorporated into models such as those of \citet{haehnelt93} and
\citet{kauffmann} in which simple parameterisations of such
cosmological evolution in accretion rate were allowed,
and it was shown that with a suitable choice of parameters the evolution in the
QSO optical luminosity function could be matched.  The physical
origin of this variation in accretion rate has not yet been determined, however,
and explanations range from a systematic depletion in available gaseous
material with cosmic epoch to long-term variations in the accretion process
itself:  these latter explanations would of course only work if black holes
were formed at high redshift and the luminosity evolution that has been observed
were a reflection of the variation in accretion history of individual
black holes.

More recently however it has become clear that optically luminous QSOs
are a short-lived phenomenon, at least compared with the Hubble time.  
Measurement of the clustering of QSOs
in the 2dF QSO Redshift Survey (2QZ: \citealt{croom05}) has shown that
their clustering amplitude does not increase with time as would be
expected if QSOs were long-lived, given that cosmic structure is expected
still to be growing at $z \sim 2$ in a Universe with a cosmological
matter density parameter $\Omega_M \sim 0.3$.  This
implies that we must look for a universal variation in accretion rate
rather than anything intrinsic to an individual accretion ``event''.

The simple picture of QSO luminosity evolution itself has now been shown to
be more complicated when active galaxies are selected at X-ray
wavelengths.
First, it now seems likely that the hard X-ray background is produced
by X-rays emitted from active galaxies, but in that case a large fraction
of those active galaxies must be highly absorbed, with equivalent
X-ray absorption column densities in neutral hydrogen $> 10^{23}$\,cm$^{-2}$
\citep{risaliti,comastri,ueda}.  These active galaxies would be likely to
be optically obscured also.
A second significant discovery from X-ray surveys however has been
that low luminosity X-ray selected active galaxies, that would be
classified as Seyfert galaxies were they optically selected, display
cosmological evolution in their space density that appears to have a maximum
at lower redshift than their high luminosity QSO counterparts
\citep{steffen,cowie,ueda,zheng,barger}.  It may be to some extent that the
identification of these weak active galaxies may be less complete at
higher redshifts, but early indications are that there is a significant
shift of the space density maximum to lower redshifts with decreasing
luminosity.  

Hence it appears that there is much still to understand both
observationally and theoretically about the evolution of active
galaxies.  Nonetheless, the basic inference that there must have been
evolution in accretion rate seems inescapable.  In this paper we
test the extent to which the growth of black holes may be linked to the
growth of galaxies, and argue that cosmological evolution in accretion rate
can be understood primarily as arising from the cosmological evolution
in the rate of accretion of matter onto galaxies.  
In section 3 we
present a new calculation of the accretion rate onto dark halos
using the extended Press-Schechter approximation.
We find that the analytic expression agrees well with both our own
and previous numerical estimation of the accretion rate calculated
from merger trees.  The analytic approach here circumvents the need
for such numerical estimation, and we also find that it provides
mathematically better behaviour than fitting functions to the
numerical results that have previously been proposed in the literature.

In section 4 we argue that coeval growth of black
holes with their associated galaxies and dark halos leads to a simple
expression for the total mass accretion rate onto all black holes. 
We calculate the expected integrated luminosity density arising from
accretion onto black holes and show that it agrees remarkably well with
observation.
When expressed as the Eddington ratio (the ratio of the actual mass
accretion rate to the rate required to attain the Eddington limiting
luminosity) it shows significant cosmological evolution almost
independent of halo mass and without depending strongly on the choice
of cosmological parameters.

\section{Coeval growth of supermassive black holes and their host galaxies and halos}
\label{sec:coevalgrowth}
One hypothesis, whose consequences will
be followed in this paper, is that the population of supermassive black
holes, that in the nearby universe inhabit the nuclei of massive
galaxies, grew coevally with their host galaxies and associated dark halos. 
Before launching into
detailed calculation, in this section we first consider the rationale
for testing this hypothesis. 

First, 
given present data on the black hole/bulge relationship, some degree of
coeval growth seems hard to avoid, at least for galaxies with measurable bulge
components.  In all massive galaxies where both a bulge mass and a
black hole mass have been measured dynamically, there is an extremely
tight correlation between the two \citep{gebhardt00, tremaine, onken}.
At lower masses, with velocity dispersion $\sigma <
100$\,km\,s$^{-1}$, the correlation is less well-established, but does
appear to extend at least down to $\sigma \sim 30$\,km\,s$^{-1}$
\citep{barth}.  When the bulge mass is determined from velocity
dispersion measurements, the cosmic scatter in the relation appears
smaller than the measurement errors, and probably is less than a
factor three \citep{tremaine,onken}.  But it is thought that galaxies
have grown hierarchically over a wide range of cosmic epoch, and with
galaxies at any observed redshift having accumulated their masses at
many different times.  In order for the tight present-day
black-hole/bulge relation to exist, the process of formation of black
hole and galaxy must have been interrelated, the simplest explanation
being that they grew in mass at the same time.

Second, the luminosity emitted by AGN is thought to arise from accretion,
and therefore is a signature of ongoing black hole growth at low redshifts.
\citet{ueda} and \citet{marconi} have shown that both the X-ray background and
the local mass density in black holes are consistent with being created by
the luminous phase of accretion visible in hard X-ray AGN surveys,
and that the bulk of that black hole growth occurs at low redshifts ($z < 3$).
The picture of supermassive black holes forming at high redshifts and then
remaining largely unchanged since then is not consistent with the
observed AGN luminosity density and the inferred local black hole mass density,
unless the radiative efficiency of the AGN luminosity we can see is unfeasibly
high ($\epsilon >> 0.1$, \citealt{marconi}).  So it appears that black holes
have been continuing to form during the cosmic epochs in which dark halos
and their associated galaxies have also continued to grow.

Previous AGN models (e.g. \citealt{kauffmann}) have assumed a relationship
between dark halo mass and black hole mass.  It is hard to see how the black
hole could know what mass of halo it is in unless there has been some causal link between
halo growth and black hole growth.  The link does not need to be direct, and it
may be that feedback processes have an important role in regulating the
black hole/halo relationship, but in effect, these models have
implicitly assumed that coeval growth has occurred.  Whether coeval growth is
{\em still} occurring in the present-day universe is another question, and one
that we shall attempt to investigate in this paper.

The growth of galaxies by hierarchical mergers implies that 
there is a contribution of mergers to black hole growth, too.
Thus we expect the {\em mass function} of black holes to be determined
by hierarchical merging, as is the mass function of galaxies.  Mergers
of black holes cannot change the integrated irreducible mass in black
holes, however, so the {\em integrated mass} in black holes should
depend on the total amount of matter accreted and not on the merger
history.  The importance of mergers does not violate the argument
originally due to \citet{soltan} that black holes in the local
universe are produced by accretion at earlier cosmic epochs.  The
growth of supermassive black holes through mergers does in
principle allow high black hole masses to be attained at rates faster
than the Salpeter rate without violating the Eddington limit, which
may help us to understand how supermassive black holes can exist
at high redshift, $z \sim 6$ (e.g. \citealt{willott}).

In practice we might expect the phases of significant mass accretion
in the life of a black hole to be correlated with periods of galaxy
merging: it has long been suggested that mergers may drive matter into
galaxy centres and trigger phases of black hole accretion
(e.g. \citealt{barnes92}).  This picture creates an attractive link
between galaxy mergers and prompt AGN activation but is not required
in what follows.

\section{The mean accretion rate onto dark matter halos}

\subsection{Structure growth from the Press-Schechter approximation}

If the matter content of the universe is dominated by cold dark matter,
then dark matter halos associated with galaxies form hierarchically.  
\citet{bond} have shown that a more
rigorous treatment of the work of \citet{ps}
(``extended Press-Schechter'', hereafter EPS)
can be used to obtain information about the build-up of structure,
and the resulting evolving mass functions agree with the
results of N-body simulations, especially if some additional
modification of EPS theory is allowed \citep{sheth99, sheth01, sheth02}.  

In this picture of hierarchical galaxy formation, at any moment in
time, any given overdensity is increasing in mass through the
process of accretion of matter.  There has been much discussion in
the literature about whether one can use EPS theory to further
analyse this accretion in terms of merger events \citep{lc93,
lc94,cole,benson} but we are not interested here in
attempting such a detailed view of the build-up of galaxies.  
In previous papers \citep{percival99, percival00}
the formation rate of halos within EPS theory has been calculated,
and hence if the space density of halos is also known this may be
converted into a mean accretion rate onto halos as a function of
their mass and redshift, measuring the 
overall accretion of matter, not broken down into individual events.

In those papers it was shown that a joint distribution function 
in mass $M$ and
cosmic time $t$ of new overdensities, $F(M,t)$, could be defined that
is related to the mass function, $F(M|t)$,
by the relation
\be
  F\left(M,t\right) dM dt = F\left(M|t\right) 
    \left|\frac{d\delta_c}{dt}\right| dM dt,
\ee
where $\delta_c(z)$ is the linear-theory overdensity at redshift $z=0$ that
would be required for an overdensity to have collapsed at some redshift $z>0$.
For an Einstein-deSitter universe 
\be
\left|\frac{d\delta_c}{dt}\right| \simeq 1.686 H_0 \left(1+z\right)^{2.5}.
\ee
In Appendix A we discuss alternative methods of calculating
$|d\delta_c/dt|$ for other choices of cosmological parameters, and describe
an analytic approximation based on the linear growth factor of cosmological
perturbations.

However, the normalisation of the joint distribution $F(M,t)$ is problematic,
as discussed by \citet{percival00}. In
Section~\ref{sec:analytic} we follow a different approach and
calculate the mean accretion rate analytically from the EPS
conditional mass function. This is complemented in
Section~\ref{sec:montecarlo} by a numerical derivation of the rate
from Monte-Carlo realisations of the EPS process.

\subsection{Analytic derivation of the mean halo accretion rate}  
  \label{sec:analytic}

\citet{lc93,lc94} give the conditional probability that a halo of mass
$M$ at cosmic time $t$ was previously a halo of lower mass $M'$ at earlier
time $t'$:
\begin{eqnarray}
\nonumber
  \lefteqn{\frac{df}{dM'}\left(M',t' | M,t \right) = }\\
  & & \frac{\Delta\delta_{c}}{\sqrt{2\pi}\left(\Delta\sigma^2\right)^{3/2}}
  \left| \frac{d\sigma^2}{dM'} \right|
  \exp\left[-\frac{\left(\Delta\delta_{c}\right)^2}{2\Delta\sigma^2}\right],
  \label{eqn:condmass}
\end{eqnarray}
where $\Delta\delta_{c} = \delta_{c}(t')-\delta_{c}(t)$, $\delta_{c}(t') > \delta_{c}(t)$,
and where  $\Delta\sigma^2 = \sigma'^2 - \sigma^2$ with
$\sigma'^2$ being the variance on the mass scale $M'$, 
$\sigma^2$ the variance on the mass scale $M$
and $\sigma'^2 > \sigma^2$ for $M' < M$.

Hence the expectation value for the halo's increase in mass  $\Delta M
\equiv M - M'$ is
\begin{eqnarray}
\nonumber
  \lefteqn{\left\langle \Delta M \right\rangle =} \\
   & & \hspace*{-5mm} \int_0^{M} \frac{\left(M-M'\right) 
  \Delta\delta_{c}}{\sqrt{2\pi}\left(\Delta\sigma^2\right)^{3/2}}
  \left| \frac{d\sigma^2}{dM'} \right|
  \exp\left[-\frac{\left(\Delta\delta_{c}\right)^2}{2\left(\Delta\sigma^2\right)}\right]dM'.
\end{eqnarray}
Making the
substitution $y^2 = 1/\Delta\sigma^2$ this can be rearranged to give
\begin{eqnarray}
\nonumber
  \lefteqn{\left\langle \Delta M \right\rangle =} \\
   & & \hspace*{-5mm} M \Delta\delta_{c} \sqrt{\frac{2}{\pi}}
  \int_0^{\infty} \left(1-\frac{M'(y)}{M}\right)
  \exp\left[-y^2\left(\Delta\delta_{c}\right)^2/2\right] dy.
  \label{eqn:exact1}
\end{eqnarray}
As $y \rightarrow \infty$, $M'(y) \rightarrow M$ and
$\left(1-\frac{M'(y)}{M}\right) \rightarrow 0$. Hence we can make
$\Delta\delta_{c}$ sufficiently small that the exponential term can be
ignored in this integral, and in the limit $\Delta t = (t - t')
\rightarrow 0$, $\Delta\delta_{c} \rightarrow 0$ and
\begin{eqnarray}
  \left\langle \Delta M \right\rangle & \rightarrow &
  M \Delta\delta_{c} \sqrt{\frac{2}{\pi}}
  \int_0^{\infty} \left(1-\frac{M'(y)}{M}\right) dy\\
& \rightarrow &
  M \frac{\Delta\delta_{c}}{\sqrt{2\pi}}
  \int_0^{\infty} \left(1-\frac{M'}{M}\right) 
  \frac{d\Delta\sigma^2}{\left(\Delta\sigma^2\right)^{3/2}},
\end{eqnarray}
where $\Delta\sigma^2 \rightarrow \infty$ as $M' \rightarrow 0$
for CDM power spectra 
and $\Delta\sigma^2 \rightarrow 0$ as $M' \rightarrow M$.
The kernel of this integral was derived by \citet{cole} but
those authors were interested in considering the distribution
of merger events rather than integrating to find the overall
mean accretion rate, which forms the basis of this paper.

Hence we can write the mean rate of mass accretion as
\be
  \left\langle \dot{M} \right \rangle =
  \lim_{\Delta\delta_{c} \rightarrow 0} \frac{\left\langle \Delta M \right\rangle}{\Delta\delta_{c}} 
  \left| \frac{d\delta_{c}}{dt} \right |
  = M \left| \frac{d\delta_{c}}{dt} \right | f\left( M \right),
\label{eqn:meanrate}
\ee
where
\be
  f\left( M \right) = \frac{1}{\sqrt{2\pi}}
  \int_0^{\infty} \left(1-\frac{M'}{M}\right) 
  \frac{d\Delta\sigma^2}{\left(\Delta\sigma^2\right)^{3/2}}.
\ee

Integrating by parts we find
\begin{eqnarray}
\nonumber
  f\left( M \right) & = &
  \sqrt{\frac{2}{\pi}}
  \left[ \left(1-\frac{M'}{M}\right) \left(\Delta\sigma^2\right)^{-1/2} \right]_{M'=0}^{M'=M}
  + \\
& & 
  \sqrt{\frac{2}{\pi}}
  \int_{0}^{M} \left(\Delta\sigma^2\right)^{-1/2}
  \frac{dM'}{M}.
  \end{eqnarray}
The first term is zero provided $\left| \frac{d\sigma^2}{dM'} \right|_{M'=M} > 0$, 
leaving
\be
  f\left( M \right) =
  \sqrt{\frac{2}{\pi}} \frac{1}{M}
  \int_{0}^{M} \left(\Delta\sigma^2\right)^{-1/2} dM'.
\ee

It should be noted that this particular definition of mass accretion rate
is not the only one possible.  One could also calculate the
accretion rate by considering the future accretion of halos and
taking the limit of small time intervals:  this produces a different
analytic form, involving an integration over masses larger then the 
mass being considered, and highlights that the mass accretion process
for a mass $M$ at time $t$ is discontinuous within EPS theory. 
In section \ref{sec:montecarlo}
we also consider the relationship of this ``instantaneous'' mass accretion rate
to the process of the build-up of halos on longer timescales:
the mean accretion rate is a function of the time interval being
considered, as may be seen from inspection of Eq.~\ref{eqn:exact1}.  

\begin{figure}
  \resizebox{8cm}{!}{ 
  \rotatebox{-90}{
  \includegraphics{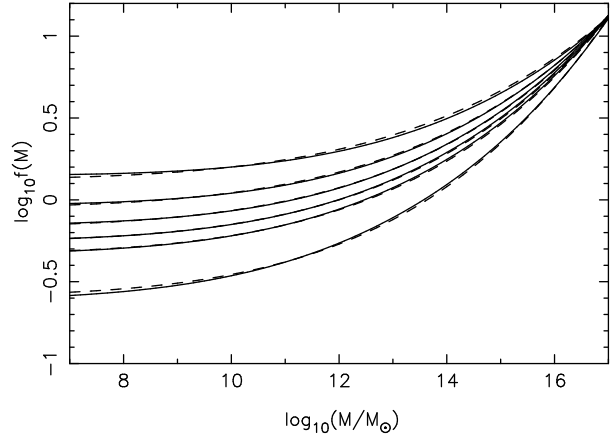}}}
  \caption{
The variation of the function $f(M)$ with log$_{10}$(halo mass) as a function
of the power-spectrum shape parameter $\Gamma$, with $h=0.7$
and $\sigma_8 = 0.74$.  
Curves from top to bottom have
$\Gamma = 0.07, 0.14, 0.21, 0.28, 0.35, 0.7$; solid curves show the calculated 
function, dashed curves show the fitting function, equation\,\ref{eqn:fitting}.
}
\label{fmplot}
\end{figure}

The function $f(M)$ may be evaluated numerically for an assumed matter
power spectrum.  For cold dark matter power spectra it depends only
weakly on mass and varies little with shape of the power spectrum.  It
varies inversely with the normalisation parameter $\sigma_8$.
Fig.\,\ref{fmplot} shows the variation of $f(M)$ with halo mass for a
$\Lambda$CDM power spectrum approximated as in \citet{ebw92} for
a range of values of the shape parameter $\Gamma \equiv \Omega_M h$
and with fixed $\sigma_8 = 0.74$, $h=0.7$.  
The relationship between variance $\sigma^2$ and mass was
calculated assuming a spherical top-hat smoothing function, although
we should note that strictly Eq.~\ref{eqn:condmass} is derived assuming
smoothing with a top-hat function in $k$-space.
At a halo mass of $10^{12.5} h^{-1} \msun$, appropriate for the halo masses of
luminous QSOs \citep{croom05}, $f(M)$ varies only by a factor 2 over the
range $0.1 < \Omega_M < 0.4$ for $h=0.7$.  The variation with mass is
also weak for masses in the range appropriate for massive galaxies:  
at $\Omega_M = 0.3, h=0.7$, $f(M)$ varies by a factor 2 over the
range $10 < \log_{10}(M/\msun) < 13$.

For convenience we provide a fitting formula for $f(M)$.  Over the range of
$\Gamma$ and mass shown in Fig.\,\ref{fmplot}, we adopt the relation
\begin{equation}
\log_{10}\left(f\left[M\right]\right) =
A\left(\Gamma\right) + \left(B - A\left(\Gamma\right)\right)
\left[\log_{10}\left(
\frac{(h/0.7)^2 M}{10^{17}\msun}\right)\right]^{\beta}
\label{eqn:fitting}
\end{equation}
with $B = 1.117$, $\beta = 4.82$ and
$A\left(\Gamma\right) = -0.7349 - 0.9808(\log_{10}\Gamma) - 0.2055(\log_{10}\Gamma)^2$.
This function fits to an r.m.s. accuracy of $1.6$\,percent over the range
$0.07 \le \Gamma \le 0.7$, or $0.1 \le \Omega_0 \le 1$ for $h=0.7$,
and is shown by dashed lines in Fig.\,\ref{fmplot}.  
For $\sigma_8 = 0.84$ the best fitting function parameters are
$B = 1.072$, $\beta = 4.57$ and
$A\left(\Gamma\right) = -0.8665 - 1.002(\log_{10}\Gamma) - 0.2093(\log_{10}\Gamma)^2$.
This formula
may be used together with the 
analytic approximation to $|d\delta_c/dt|$ described in Appendix\,A
to calculate the mean cosmological accretion rate for a wide range of
cosmological parameters.  
It is worth noting that the accretion rate has a simple dependence
on mass and redshift, and either weak or at least linear dependence on
cosmological parameters.  

\subsection{Numerical calculation of the mean halo accretion rate}  
  \label{sec:montecarlo}

\begin{figure}
  \resizebox{8cm}{!}{ 
  \rotatebox{0}{
  \includegraphics{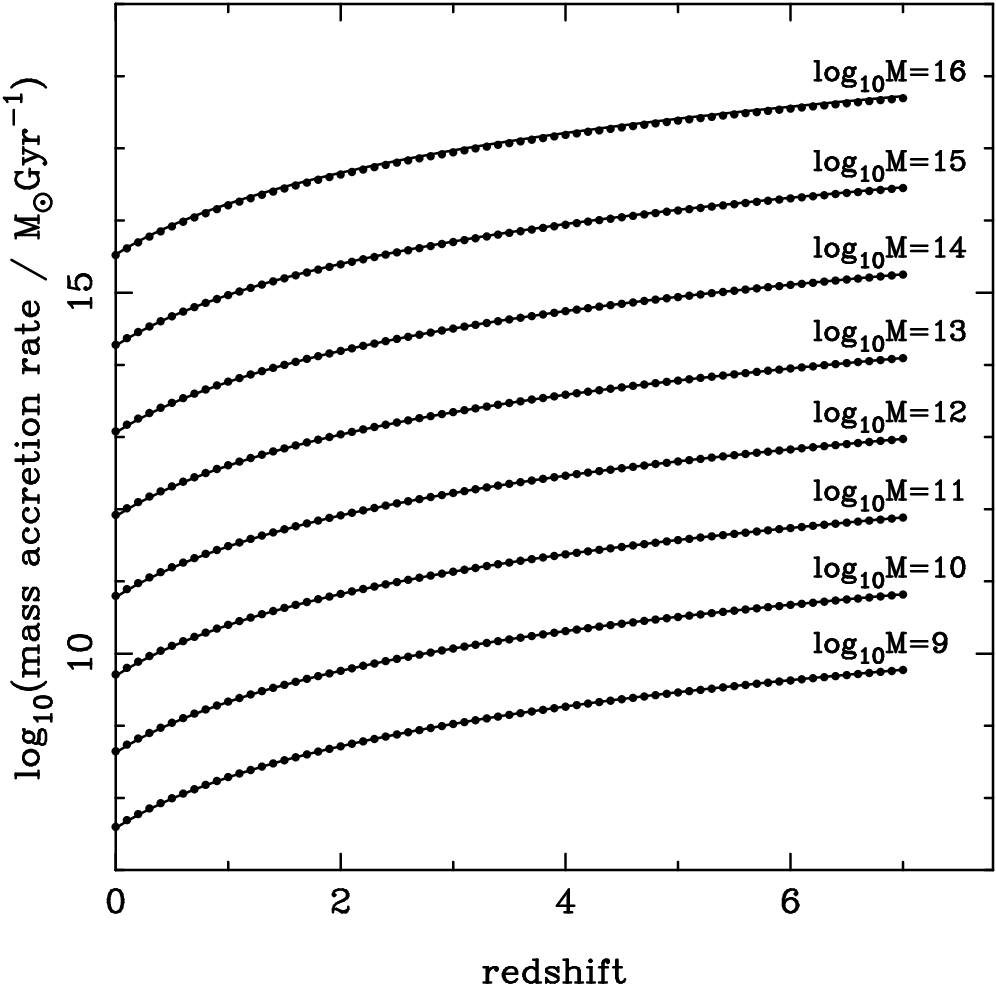}}}
  \caption{Comparison of the mean mass accretion rate calculated
  analytically (solid curves) and measured from Monte-Carlo
  realisations (solid symbols), shown as a function of redshift for a
  variety of masses.}
\label{fig:PS1}
\end{figure}

To validate the derivation presented in Section~\ref{sec:analytic}, we
now compare the analytic instantaneous rate with the rate calculated
from Monte-Carlo realisations of the EPS process. In EPS theory, the
overdensity recovered if a realisation of the density field is
smoothed on different scales around a fixed point can be matched to a
one-dimensional Brownian random walk. The collapsed mass at a given
time is given by the trajectory's first upcrossing of a barrier at the
critical overdensity corresponding to the chosen time. Because we can
only calculate finite representations of these trajectories, there are
a number of numerical issues that complicate their use. In particular
\begin{itemize}
   \item the discrete nature of the numerical trajectories means that
   we can miss upcrossings of a barrier between steps in the
   realisations. If the step size is chosen to be too large, then this
   effect becomes particularly apparent for the first step in each
   trajectory, where most upcrossings are expected.
   \item the number of steps must be finite, so we will always miss
   upcrossings at high $\sigma^2$ (small mass).
\end{itemize}

By appropriate choice of the step size, we can minimize these
problems. We have created a sample of $10^6$ trajectories each
consisting of $2^{13}$ steps of varying size in $\sigma^2$.  A varying
step size was chosen to more closely match the expected upcrossing
distribution and allow a larger range in $\sigma^2$ to be probed. The
barrier height was set at $4$ times the initial step size to minimise
problems owing to missed upcrossing in the first few steps of the
trajectories. Because the trajectories can be arbitrarily scaled in
$\delta_{c}^2/\sigma^2$, we can use this single set of trajectories
(and associated upcrossings) to determine the mass accretion rate for
any halo mass and epoch. For the masses and times chosen, we have
determined that the problems owing to the finite extent of the
trajectories are negligible. By comparing with the expected
conditional mass function (Eq.~\ref{eqn:condmass}), we have also
optimised the step size to reduce problems owing to the discreteness of
the realisations.

\begin{figure}
  \resizebox{8cm}{!}{ 
  \rotatebox{0}{
  \includegraphics{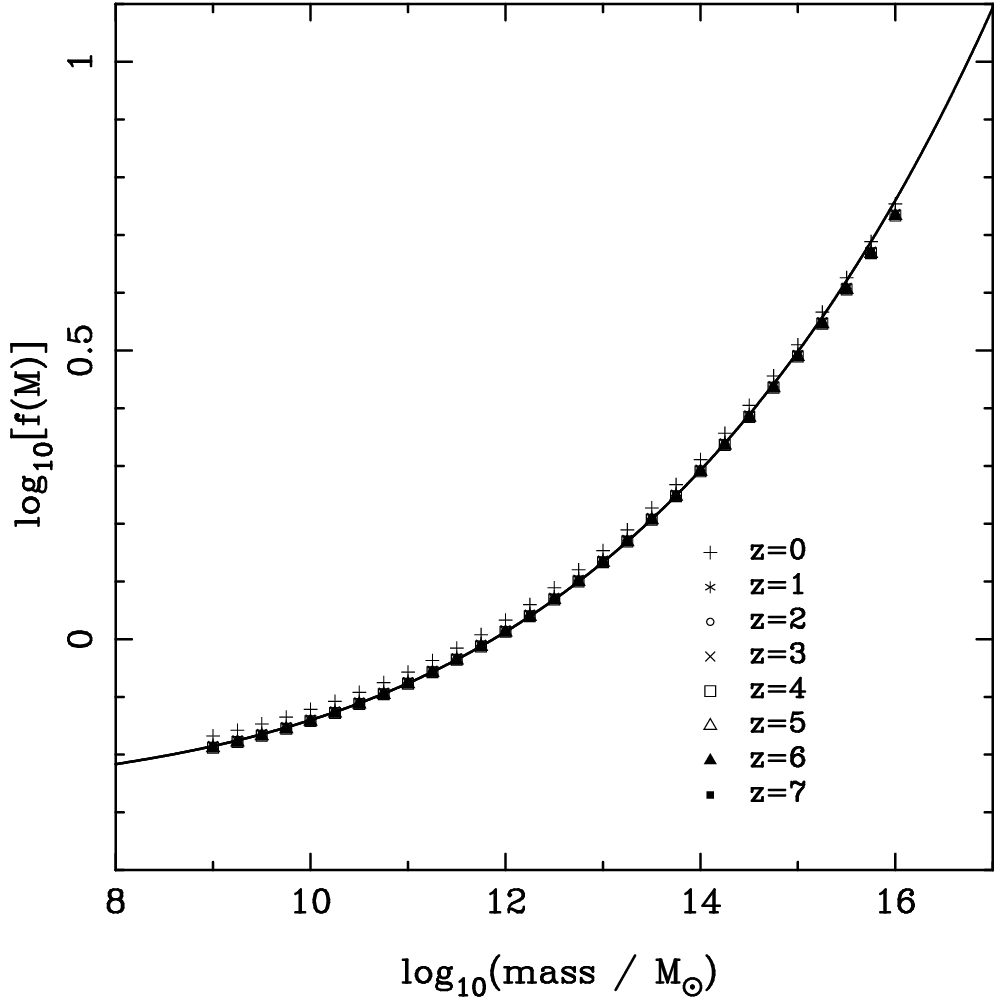}}}
  \caption{
Comparison of $f(M)$ (solid curve) with that
deduced from the Monte-Carlo realisations by factoring out the linear dependence
on $M$ and $d\delta_{c}/dt$.
}
\label{fig:PS2}
\end{figure}

Fig.\,\ref{fig:PS1} shows the result from the Monte-Carlo measurements,
calculated from the average mass accreted within
$\Delta\delta_{c}=0.032$, corresponding to $0.0035$\,Gyr at $z=7$ and
$0.5$\,Gyr at $z=0$. Decreasing $\Delta\delta_{c}$ further does
not significantly change the recovered rate. The agreement with the
analytic instantaneous rate is extremely close. Any differences 
may be revealed by 
comparing $f(M)$ with the equivalent quantity from the Monte-Carlo
realisations by factoring out the analytic linear dependence on mass
and $d\delta_{c}/dt$ (Fig\,\ref{fig:PS2}). The maximum difference
between the analytic and numerical results is $<5 $\% and is
likely to be caused by remaining problems owing to the discrete nature
of the numerical trajectories.  Note that 
the curves in Fig.\,\ref{fig:PS1} show the immediate mass accretion rate
onto a halo which has a specified mass at a specified redshift,
not how the mean accretion rate varies with time for
an individual halo.

This comparison has shown that the results obtained from numerical
realisations of the Press-Schechter process
are indeed consistent with the analytic results that we have obtained.  
This type of realisation forms the basis of the commonly-used ``merger trees''
and in the next section we compare our results with earlier numerical
attempts to understand the growth of dark matter halos.

\subsection{Comparison with previous numerical calculations}
\citet{vandenbosch} has calculated the average mass accretion
histories of dark matter haloes based on the extended Press-Schechter
formalism and the $N-$branch merger-tree construction algorithm of
\citet{somerville}.
He provides a fit to the numerically-obtained mass accretion histories by
fitting a function of the form
\begin{equation}
\log\left\langle \frac{M(z)}{M_0}\right\rangle = -0.301\left[\frac{\log (1+z)}{\log (1+z_f)}\right]^\nu ,
\label{eqn:vdbfit}
\end{equation}
where $M(z)$ is a halo's mass at redshift $z$,
$M_0$ is its present-day mass, and $\nu$ and $z_f$ are non-independent parameters.
The best-fit set of parameter values is
$\nu = 1.362 + 1.858\log
(1+z_f)-0.032\log \left[M_0/(10^{11}h^{-1}\msun)\right]$.  $z_f$ is
obtained from the root of equation:
\begin{equation}
\delta_c(z_f) = \delta_c(0) + 0.477 \sqrt{2\left[\sigma^2(f_{i}M_0)-\sigma^2(M_0)\right]}
\end{equation}
where $f_i$ is set to a value $f_i=0.254$.
\citet{vandenbosch} recommends differentiating $M(z)$ in order to find the mean mass
accretion rate.  As discussed by \citet{vandenbosch}, construction of such accretion
histories numerically is not trivial, and depends on the definition of a halo's 
progenitor within a merger tree.  However, it will be interesting to compare the
results of the analytic result presented here with these earlier numerical attempts.

\begin{figure}
  \resizebox{8cm}{!}{ 
  \rotatebox{-90}{
\includegraphics{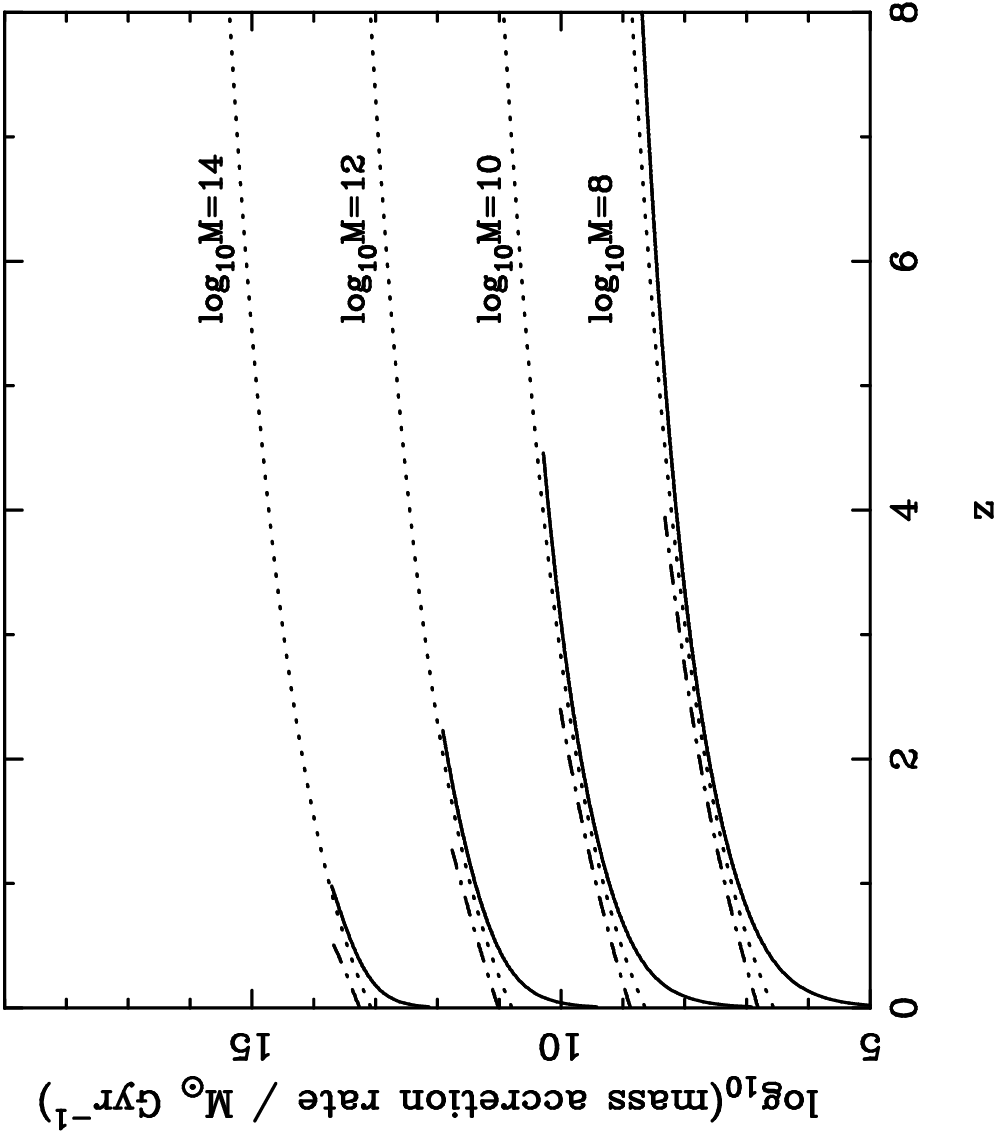}}}
\caption{
Comparison of the mean accretion rate from this paper (dotted lines)
with the two alternative numerical fits proposed by \citet{vandenbosch}
(solid lines) and by \citet{wechsler} with parameter values estimated
by \citet{vandenbosch} (dot-dashed lines).  
See the text for more details of the
calculation of these curves, which are shown for values of
$\log_{10}(M(z)/\msun)=8,10,12,14$.  The \citet{vandenbosch} curves
becomes unphysical at low redshift and depart significantly from the
analytic results.  Both numerical fits become unphysical at high
redshift and are truncated outside the region of validity, as discussed
in the text.
}
\label{fig:MAH}
\end{figure}

\citet{wechsler} have also discussed the growth of halos in merger trees, and 
find a reasonable parameterisation of the accretion histories of 
halos to have the form:
\begin{equation}
M(z) = M_0 e^{-\alpha z}.
\label{eqn:wechsler}
\end{equation}
where $\alpha$ is a parameter.
In fact, comparison with equation\,\ref{eqn:meanrate} suggests this is a good 
approximating function for \citet{wechsler} to have chosen. From 
equation\,\ref{eqn:meanrate} we find 
\begin{equation}
\frac{dM(z)}{dz} = -M f\left( M \right) \frac{d\delta_{c}}{dz}.
\end{equation}
In the case of an Einstein-de-Sitter universe $d\delta_{c}/dz = 1.686$ and a
solution for $M(z)$ of the form of equation\,\ref{eqn:wechsler} immediately follows
if we ignore the slight redshift dependence introduced by the $f(M(z))$ term.
This solution will not be correct for other cosmologies, but should still
qualitatively be of the correct form.
\citet{vandenbosch} has related the parameter $\alpha$ to the previous
parameter $z_f$ by the relation
$\alpha = \left(z_f/1.43\right)^{-1.05}$.

In Fig.\ref{fig:MAH}
we compare our accretion rate with those obtained from these two
alternative fits to the numerical mass accretion histories. 
The solid line shows the mean accretion rate obtained by
differentiating equation\,\ref{eqn:vdbfit} and the dot-dashed line
from differentiating equation\,\ref{eqn:wechsler}, both with the
prescriptions for the free parameters given by \citet{vandenbosch}
that are quoted above.  
The dotted line shows the mass accretion rate from equation\,\ref{eqn:meanrate}
(note that, as in Fig.\,\ref{fig:PS1}, 
each curve corresponds to a fixed value of mass at each redshift:
i.e. these are curves of constant $M(z)$ and not constant $M_0$).
The comparison is made for several values of halo mass:
$\log_{10}\left(M(z)/\msun\right)=8,10,12,14$. 
There is excellent
agreement between the three curves over most of the range of redshifts.
However, at the lowest redshifts the \citet{vandenbosch} accretion
rate shows significant departures, and it is straightforward to see
that this arises from the choice of fitting function in equation\,\ref{eqn:vdbfit}:
in the limit $z \rightarrow 0$, $dM/dt \rightarrow 0$ if calculated from
that equation, for parameter values $\nu > 1$.  Such behaviour is unphysical,
it implies that all halos at the present day have stopped growing.
This is purely an artefact of the choice of fitting function and 
leads to significant departures from the analytic calculation for $z<0.5$.
The result from the \citet{wechsler} form consistently results in
slightly higher values of accretion rate, but predicts the same
behaviour as the analytic calculation at redshifts approaching zero.

We also find a further limitation of the numerical prescriptions,
which is that the calculated growth histories become unphysical
at high masses and redshifts, with problems arising when the present-day
mass $M_0\gtrsim 10^{15}\msun$.  The problem is that the $M(z)$
functions for differing masses cross at high redshift, such that a high mass 
halo might be predicted to have had a lower-mass progenitor than a lower mass
halo.  For example, the main progenitor
mass of a $M_0=10^{16}\msun$ halo is predicted by equation\,\ref{eqn:vdbfit}
to be smaller than that of a $M_0=10^{14}\msun$ halo at redshifts $z\gtrsim 2.5$. 
We roughly estimate that the \citet{vandenbosch} fit can be used up to
$(\log_{10}(M_0/\msun),z)=\{(14.5,1),(13.3,2),(11.7,4),(10.8,6),(10.2,8.)\}$,
where the pairs of numbers denote the locus on the $M_0, z$ plane beyond which
the prescription becomes unreliable.
The same problem is found when using the parameterisation 
of equation\,\ref{eqn:wechsler}, 
with the range of valid redshift and present-day halo mass moved to lower values:
$(\log_{10}(M_0/\msun),z)=\{(13.3,1),(11.8,2.),(9.8,4),(8.3,6),(7.7,7)\}$.
Given the relationship between the \citet{wechsler} and analytic functions, it
can be seen that the problem in this case arises from the prescription for
the value of $\alpha$
rather than being an intrinsic problem with the choice of numerical function.

Overall we conclude that the analytic and previous numerical results are in 
reasonable agreement in the regions of mass and redshift space where the numerical
parameterisations are reliable, but that there exist significant regions of
that space where the previous numerical fits become unusable.

\subsection{How good is extended Press-Schechter theory?}
So far, both the analytic calculation and the numerical realisations have relied
on the EPS formalism.  This formalism effectively amounts to
being a linearised approximation to the true process of non-linear growth of
structure, and {\em a priori} we should have no right to expect that this will
be an accurate approximation.  How can we be confident that the work presented
here has relevance to the real process of halo growth?  
There have been many comparisons of the calculation
of the halo mass function predicted by EPS with that obtained from N-body dark
matter simulations and, with some caveats, good agreement is found.  \citet{jenkins}
show the level of agreement when compared with Virgo consortium simulations.
The mass functions agree to a factor 1.6 over a wide range of halo mass.  
In this paper we investigate the rate of growth of dark halos: if the EPS growth
rate were significantly wrong it would be hard to then reproduce a good estimate
of the mass function, and it 
seems likely therefore that the calculations presented here are correct when
compared with dark-matter-only simulations to this level of accuracy.

\citet{sheth99} have shown, however, that the EPS mass
function may be modified to produce even closer agreement with the N-body
simulations. This work has been developed further by \citet{sheth01} 
and \citet{sheth02} and confirmed in the \citet{jenkins} simulations.  The principal
modification is to reduce the critical density for collapse, $\delta_{c}$, by
a factor $\sqrt{a}$ where $a$ is determined from comparison with the simulations
to have a value $a \simeq 0.7$.  Insertion of this factor into 
equation\,\ref{eqn:condmass} would simply imply that $f(M)$ is also reduced 
by a factor $\sqrt{a} \simeq 0.84$.  This causes a slight shift in the curves 
plotted here but otherwise has no effect on the redshift or mass dependence.

We note that the merger trees of \citet{vandenbosch} have also been compared
with N-body simulations, and reasonable agreement was found there too, albeit
with a tendency for halos to form at higher redshifts which could also be
reproduced by introduction of the factor $\sqrt{a}$.  In the case of the 
\citet{vandenbosch} simulations the best-fitting value 
for $\sqrt{a}$ varied over the range
$0.82 < \sqrt{a} < 0.94$, depending on halo mass.
Hence it may be that there is some
systematic and possibly mass-dependent departure of the EPS accretion rate
from the ``true'' (dark-matter N-body) rate, but most likely at a 
level $<20$\%\ in accretion rate.
Finally, we note that 
\citet{benson} have also argued that the distribution of mass discontinuities
in EPS does not allow a self-consistent interpretation of these as merger
events. This casts doubt on one of the fundamental assumptions of EPS-based
merger trees, but should not affect the calculation of the mean accretion rate,
which does not rely on that assumption.

\section{The accretion history of supermassive black holes}

\subsection{Pure coeval evolution}
\label{sec:pce}
We now move on to consider what effect the evolution in mass accretion
might have on the ultimate accretion onto a supermassive black hole
lying at the heart of a massive galaxy.  In principle, understanding 
not only the formation of galaxies within dark halos but also
the accretion of baryonic matter onto a central black hole requires
complex and poorly-constrained physics.  However, we could make a very
simple assumption for the net effect of these complex processes, and see
how well the predictions compare with observation.  The assumption we
shall make is that, {\em on average}, 
the relative growth rates of central black holes
track the growth rates of their associated dark halos (section\,\ref{sec:coevalgrowth}).
That is,
\begin{equation}
\left\langle\frac{1}{M_{\rm BH}}\frac{dM_{\rm BH}}{dt}\right\rangle \simeq
\left\langle\frac{1}{M_{\rm H}}\frac{dM_{\rm H}}{dt}\right\rangle =
\left\langle f\left( M_{\rm H} \right)\right\rangle \left| \frac{d\delta_{c}}{dt} \right |,
\label{eqn:coeval}
\end{equation}
where $M_{\rm H}$ is a dark halo's mass, $M_{\rm BH}$ a black hole's mass,
and where $\left\langle\right\rangle$ denotes averaging over an
ensemble of halos and black holes.  
Note also that we haven't needed to
specify what fraction of mass ends up accreting onto a black hole -
that calculation would indeed need some detailed physics - the 
hypothesis simply assumes that the ensemble-averaged
relative growth rates are the same.  We shall term this simple
hypothesis ``Pure Coeval Evolution'' (PCE).

A simple extension of this hypothesis would allow some non-linear evolution
of the relation between black holes and dark halos.  Equation\,\ref{eqn:coeval}
could be modified by a factor $\alpha$ such that
\begin{equation}
\left\langle\frac{1}{M_{\rm BH}}\frac{dM_{\rm BH}}{dt}\right\rangle \simeq
\alpha
\left\langle\frac{1}{M_{\rm H}}\frac{dM_{\rm H}}{dt}\right\rangle =
\alpha 
\left\langle f\left( M_{\rm H} \right)\right\rangle \left| \frac{d\delta_{c}}{dt} \right |,
\label{eqn:nonlinear}
\end{equation}
which would lead to a relation of the form 
$M_{\rm BH} \propto M_{\rm H}^{\alpha}$ as inferred by
\citet{ferrarese}.  We can see that the effect of such a non-linear term
would be to modify the accretion rate calculations only by a constant
factor.

\subsection{The AGN luminosity density and its evolution}
\label{sec:lumdensity}
In this section we calculate 
the observationally-determined AGN luminosity density and compare
with the value predicted by PCE.
We define the integrated luminosity density to be the 
luminosity emitted during the black hole accretion process per 
comoving cubic Mpc, summing over AGN of all luminosities, as previously
calculated by a number of authors (e.g. \citealt{marconi}).
The quantity predicted by the PCE hypothesis is the bolometric
luminosity density, so we will need to 
apply bolometric corrections to observed quantities.

\subsubsection{The bolometric luminosity density derived from X-ray surveys} 
We shall first estimate the bolometric luminosity density 
from the \citet{ueda} AGN
luminosity function.  It has recently been recognised that hard X-ray
surveys currently provide the most complete way of selecting AGN of
all types, and \citet{ueda} incorporate into their analysis correction
for X-ray obscuration in the detected AGN (although if there is a 
population of extremely absorbed, Compton-thick, 
AGN these will be missing from \citealt{ueda}'s census - see below).

Fig.\,\ref{fig:fig5} shows the bolometric luminosity density
$\rho_{\rm BOL}$
derived from the \citet{ueda} LDDE model at $2-10$\,keV 
integrated over the absorption-corrected luminosity range
$10^{40} < L_X < 10^{48}$\,erg\,s$^{-1}$.  The X-ray luminosity density
has been converted to a bolometric luminosity density 
and a small correction for missing Compton-thick AGN has also been
applied, as described below.  The luminosity density is not sensitive
to the precise limits of integration: increasing the lower limit to
$L_X > 10^{42}$\,erg\,s$^{-1}$ decreases the $z=0$ luminosity density
only by 0.08\,dex without changing the maximum value (to 0.01\,dex)
or its redshift.
Decreasing the upper limit to $L_X < 10^{46}$\,erg\,s$^{-1}$ 
has no effect to within 0.01\,dex.  Most of the bolometric luminosity
density is produced by AGN with $L_X \sim 10^{44}$\,erg\,s$^{-1}$.

Also shown in Fig.\,\ref{fig:fig5} are estimates of the
uncertainty in $\rho_{\rm BOL}$.  
To calculate the uncertainty, we first refit the
binned data shown in Fig.\,11 of \citet{ueda}, kindly provided by Y.\,Ueda,
with the same six-parameter model (see \citealt{ueda} for details of the 
model), but with the normalisation parameterised instead as the integrated
luminosity density.  We then estimate confidence intervals on the
integrated luminosity density for each of the five redshift slices 
(0.015-0.2,0.2-0.4,0.4-0.8,0.8-1.6,1.6-3.0) in turn,
marginalising over the remaining five parameters of the fit.  This
procedure thus includes the full statistical uncertainty in the data
and includes an element of uncertainty arising from the fact that to cover
the full luminosity range requires some extrapolation of the model,
although with the caveat that only the set of
model functions that may be parameterised by the six-parameter function
are allowed.  The best-fit values for each redshift slice
and the deduced uncertainties are shown in Fig.\,\ref{fig:fig5}.
Some points deviate somewhat from the nominal curve deduced from
\citet{ueda}'s best fit: this is likely because their best fit was obtained
from a maximum likelihood fit to unbinned data whereas our points and
errors were evaluated from fits to the binned data.  We nonetheless
expect that the
size of the error bars should also be a good indication of the size of the
68\,percent confidence region for the likelihood-determined
function.

\begin{figure}
  \resizebox{8cm}{!}{ 
  \rotatebox{270}{
  \includegraphics{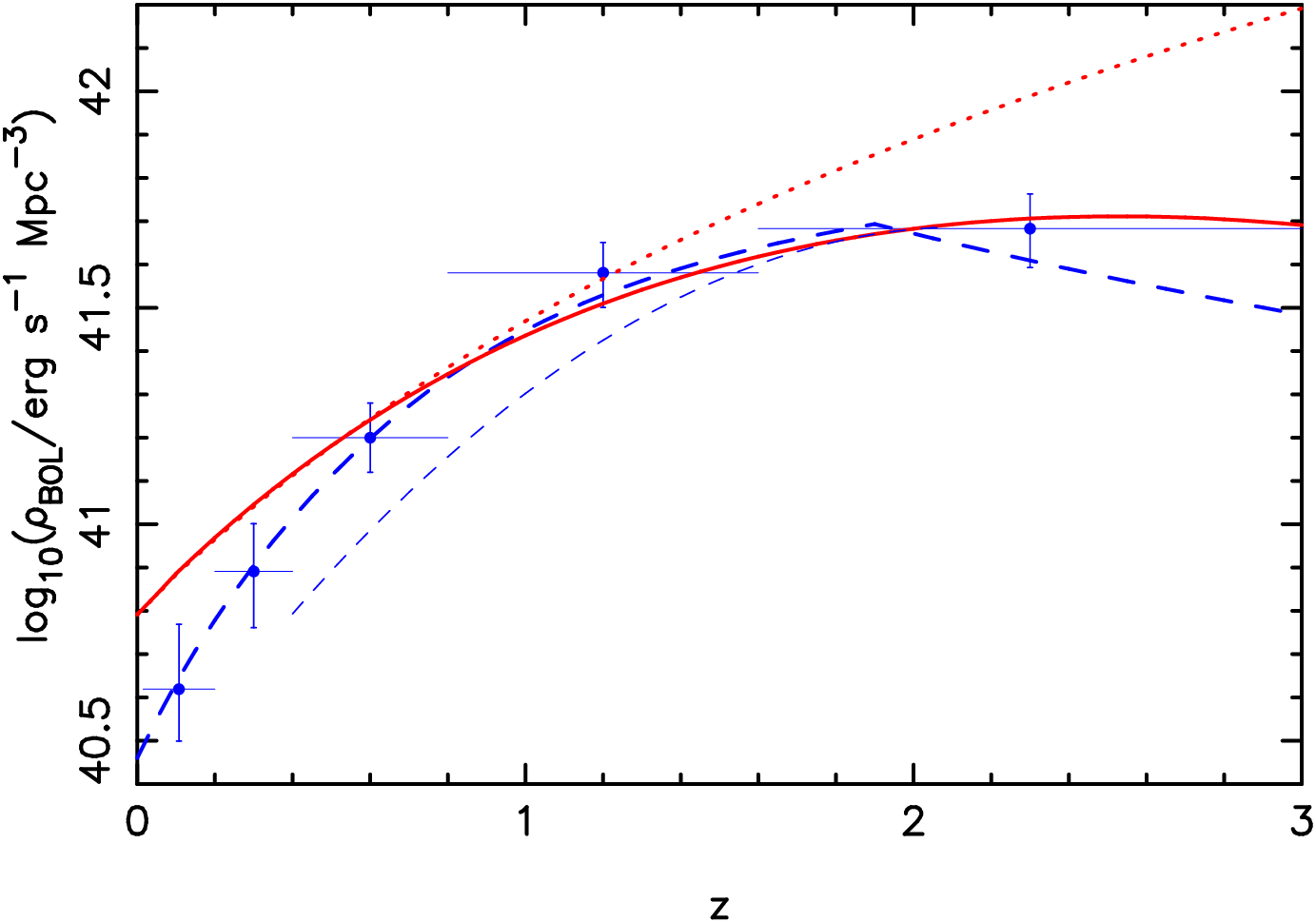}
  }}
  \caption{
The bolometric luminosity density deduced from the best-fit model of
\citet{ueda}, integrating over the range 
$10^{40}<L_X<10^{48}$\,erg\,s$^{-1}$ and applying the bolometric
correction of \citet{marconi} and correction for Compton-thick
AGN of \citet{ueda} (see text)
(dashed curve).  Also shown are uncertainties estimated
from refitting to the binned data of \citet{ueda} (points with error
bars: see text; horizontal bars indicate the range of redshifts included
in each point).
For comparison, we show the equivalent calculation from the 
pure-luminosity-evolution fit of \citet{richards} to the optical 2SLAQ
QSO survey in the redshift range $0.4<z<2.1$ (lower dashed curve).
The luminosity density expected in PCE is shown for two cases:
(i) no evolution in the comoving black hole mass density (dotted upper
curve); (ii) evolution in the comoving black hole mass density that tracks
the evolution of massive dark halos with $M_{\rm H}>10^{11.5}\msun$
(solid curve).  
Both curves assume average radiative efficiency 
$\langle\epsilon\rangle=0.04$ (see text).  Note that
the PCE calculation predicts 
a value for the luminosity density at $z=0$, dependent only on 
$\langle\epsilon\rangle$ and cosmological
parameters, with the higher redshift evolution further dependent 
only on choice of dark halo mass:
the normalisation is not allowed to float arbitrarily.
}
\label{fig:fig5}
\end{figure}

\subsubsection{X-ray absorption and the Compton-thick fraction}
It is recognised that $2-10$\,keV X-ray fluxes must be corrected for
absorption, and \citet{ueda} derive an estimate for the 
luminosity-dependent distribution of absorption column which has been incorporated
into their analysis.  In addition, 
it is likely that some fraction of AGN are Compton-thick, that
the Compton-thick fraction is a function of intrinsic luminosity
and that even hard X-ray surveys miss these objects \citep{ueda}.
\citet{ueda} assume the (very uncertain) correction, based on the results of
\citet{risaliti}, that there are $\sim 1.6$ times as many Compton-thick AGN
as there are AGN with absorption column densities in the range
$10^{23} < N_{\rm H} < 10^{24}$\,cm$^{-2}$.  We apply the same correction
here.  Although very uncertain, the net effect is to increase the 
bolometric luminosity density by only a factor 1.48 (i.e. 0.17 dex).  
Varying the
Compton-thick fraction from this nominal value would have an equivalent
effect on the value of $\langle\epsilon\rangle$ deduced, but does not
significantly affect the conclusions of this paper.

\subsubsection{The bolometric correction}
To convert $2-10$\,keV luminosity to bolometric luminosity we use the
bolometric correction of \citet{marconi}, averaged over the distribution
of luminosities of the \citet{ueda} model as a function of redshift.
Because the luminosity function is distributed to higher luminosities
at higher redshifts, the average fraction of
the bolometric luminosity that appears in the X-ray band
varies from $0.056$ at $z=0$ to $0.023$ at $z=3$, resulting in a variation
in $\log_{10}(\rho_{\rm BOL})$ of $0.4$ over this redshift range  
arising from the change in mean bolometric correction. 
Hence the bolometric luminosity density shows stronger evolution,
and has a maximum at higher redshift,
than the $2-10$\,keV X-ray luminosity density alone.
The bolometric correction is notoriously uncertain,
and one key uncertainty is the non-linear dependence between 
X-ray and optical luminosity that leads to the strong luminosity-dependence
of the bolometric correction.  \citet{marconi} assume the relation of
\citet{vignali}, which has recently been extended to lower luminosities
by \citet{steffen06} and which now seems fairly robust.  To test this
further we also compare with the bolometric luminosity density derived
from optical QSO surveys.

\subsubsection{The bolometric luminosity density from optical QSO surveys}
Optical QSO surveys do not probe to 
sufficiently low intrinsic luminosities to pick up the significant 
population of Seyfert and type II AGN, and they are dominated by
broad-line, type I objects.
(e.g. \citealt{richards,steffen06,barger}).
However, it will be a useful test of the \citet{marconi} bolometric
correction to attempt to measure the bolometric luminosity density
from an optical QSO survey:  the X-ray and optical bolometric corrections
have a dependence on bolometric luminosity of opposite signs
\citep{marconi}, arising from the $L_{\rm X}-L_{\rm opt}$ correlation
already mentioned, so if the bolometric correction is wrong this will
be manifest in this test.

The optical QSO survey that to date has most successfully probed to
low intrinsic luminosity is the 2SLAQ survey \citep{richards}. 
In Fig.\,\ref{fig:fig5} we plot the bolometric luminosity density
obtained by integrating the pure luminosity evolution model derived
from the 2SLAQ data alone, integrating over the equivalent luminosity
range as for the X-ray determination, 
with the \citet{marconi}\ optical bolometric
correction and \citet{ueda}\ Compton-thick fraction as above.  Because
of optical selection biases the valid redshift range is restricted to
$0.4<z<2.1$ \citep{richards}.

The function shown is more sensitive to the limits of integration
than the X-ray case above, because although the large survey areas allow
optical QSO surveys 
to probe to higher AGN bolometric luminosities, the range of bolometric
luminosity covered is substantially smaller than the composite X-ray surveys
discussed above \citep{richards}.  Increasing the lower luminosity limit
to match a value $L_X > 10^{42}$\,erg\,s$^{-1}$ decreases the $z=0.4$
bolometric luminosity density by 0.19\,dex and the maximum value
by 0.14\,dex.  Again, decreasing the upper luminosity limit to match a value
$L_X < 10^{46}$\,erg\,s$^{-1}$ leads to no significant change in 
luminosity density, to 0.01\,dex.

Considering the differing selection of AGN types in optical and X-ray
surveys, 
the X-ray- and optical-derived bolometric luminosity densities show
remarkably good agreement. This demonstrates that the corrections applied
are unlikely to be too far wrong, at least over the optical-X-ray part
of the spectrum (both bolometric corrections could be missing 
components in other parts of the spectrum such as the far infrared - this 
is discussed further by \citealt{marconi}).  The differences between
the two functions are most apparent at $z\sim 0.7$, although even here they 
agree within a factor 1.6.  This difference arises from the additional
low-luminosity AGN that are found by hard X-ray surveys but not by
the large-area optical QSO surveys (see
\citealt{richards} for further discussion).  

This diagram also illustrates the limited effect of possible
``cosmic downsizing'' on the luminosity density: the optical
density is derived from a pure luminosity-evolution model, the X-ray
from a luminosity-dependent density evolution model.  We can see that the
two models are measurably different, but that any 
``cosmic downsizing'' does not dominate the redshift evolution of the
luminosity density.

\subsubsection{The predicted bolometric luminosity density}
To calculate the luminosity density expected from PCE, we assume 
an average fraction $\langle\epsilon\rangle$ of the rest energy of accreted
baryonic material is radiated.  Since the fractional accretion rate
is almost independent of mass, we can write the expected bolometric
luminosity density as
\begin{eqnarray}
\nonumber
\left\langle\rho_{\rm L}\right\rangle 
&\simeq &
c^2 \rho_{\rm BH} 
\left\langle 
\frac{\epsilon}{\left(1-\epsilon\right)M_{\rm BH}}
\frac{dM_{\rm BH}}{dt} 
\right\rangle\\
&\simeq &
c^2 \rho_{\rm BH} 
\left\langle 
\frac{\epsilon f\left(M_{\rm H}\right)}{\left(1-\epsilon\right)} 
\right\rangle
\left |
\frac{d\delta_{c}}{dt} 
\right |,
\end{eqnarray}
where $\rho_{\rm BH}$ is the cosmic black hole mass density.
For the latter quantity we follow \citet{marconi} and integrate the
black hole mass function obtained by convolving the \citet{nakamura}
luminosity function with the luminosity-$M_{\rm BH}$ relation of
\citet{marconihunt}, which integrated for $M_{\rm BH} > 10^{5}\msun$
yields a $z=0$ mass density 
$\rho_{\rm BH} = 4.8\times 10^5 \msun$\,Mpc$^{-3}$ with
an uncertainty of about 30 percent \citep{marconi}.

The result is shown in Fig.\,\ref{fig:fig5} for two cases.  The first
shows what would happen if the black hole mass density did not change with
cosmic epoch: the luminosity density would then be determined entirely by 
the invariant black hole mass density and the epoch-dependent accretion
rate, as shown by the dotted curve.  For consistency with \citet{ueda}
we assume a $\Lambda$CDM cosmology with parameters $\Omega_M=0.3$, 
$H_0=73$km\,s$^{-1}$\,Mpc$^{-3}$ (the observational values are also adjusted
to this value of $H_0$) and $\sigma_8=0.74$ \citep{spergel06}.
The remaining parameter is the average radiative efficiency 
$\langle\epsilon\rangle$ which we have
set to a value $0.04$ to approximately match data and prediction at 
$z \gtrsim 0.5$. Such a value 
is consistent both with the range of values
that might be expected for accretion onto black holes and with determinations
that compare the X-ray background with the local black hole mass function,
leading to a luminosity-function-dependent value for the radiative
efficiency specifically of luminous AGN of  
$\langle\epsilon_{\rm AGN}\rangle \sim 0.08^{+0.08}_{-0.04}$ \citep{marconi}.  
Note that this could
differ from the average efficiency for {\em all} black holes if a proportion
were growing by radiatively-inefficient accretion: taken literally the
agreement with the PCE model with $\langle\epsilon\rangle=0.04$
implies that such inefficient growth does not dominate 
the overall growth of black holes at $z \ga 0.5$, 
although some ``ADAF'' contribution is allowed.  However, 
the deduced value of $\langle\epsilon\rangle$ is degenerate with 
the rather uncertain
values of the bolometric correction and black hole mass density
and with the value of $\sigma_8$.  The local 
black hole mass density is uncertain by $\sim 30$\,percent and the value
of $\sigma_8$ currently has an uncertainty $\sim 10$\,percent 
\citep{spergel06}.  The uncertainty in the bolometric correction is not
well determined.  Also, introduction of non-linear coupling between
black hole and dark halo growth (section\,\ref{sec:pce}) 
would also modify the deduced value of $\langle\epsilon\rangle$ 
by a factor $\alpha$ (equation\,\ref{eqn:nonlinear}).
Hence we should not attach too much importance to its value, provided that it
is in the range expected for accretion onto supermassive black holes
(e.g. $\sim 0.06$ for a Schwarzschild black hole).  However, 
a significantly lower value might imply a significant contribution 
to black hole growth either from radiatively inefficient accretion
or from obscured growth.

The second case shown in Fig.\,\ref{fig:fig5}
considers also the expected evolution in the black hole
mass function: if black holes do indeed grow coevally with their host galaxies
and halos we expect their mass function to show similar evolution.  The lower
solid curve shows the result expected if the integrated black hole
mass density evolves the same way as massive dark halos with 
$M_{\rm H}>10^{11.5}\msun$.  This evolution was calculated by integrating the
\citet{sheth02} mass function above this limit.  
\footnote{Note that it would be incorrect to attempt to calculate the evolution in
$\rho_{\rm BH}(z)$ by integrating $\langle d\log M/dt\rangle$: in hierarchical
growth, mergers are important and evolution of the mass function is described
by the Press-Schechter function.}
We find that imposing
an upper limit has little effect, but that the amount of turn-down 
with increasing redshift in the 
integrated luminosity density does depend on the lower halo mass limit, with
lower masses having less high-$z$ turn-down, 
as expected from standard CDM models.

\subsubsection{Comparison of observed and predicted luminosity densities}
At $z \ga 0.5$ the simple hypothesis that black holes grow with their
parent halos seems remarkably successful.
With a fixed set of cosmological parameters,
an independently determined estimate of the local mass density in black holes
and a radiative efficiency in the expected range for AGN,
this simple model of mass build-up predicts the local 
AGN luminosity density to within the observational uncertainties.
No other tuning of the model or variation of parameters is required.

Further, the integrated luminosity density
is dominated by the contribution from AGN around the ``break'' in the
luminosity function, with $L_X \sim 10^{44}$\,erg\,s$^{-1}$:  AGN of this
luminosity typically have black hole masses around $10^8\msun$ \citep{fine}
and QSOs of this luminosity in the 2dF QSO Redshift Survey (2QZ) have 
dark halo masses independently estimated from their clustering bias of
$M_{\rm H} \sim 10^{12.5}h^{-1}\msun$ \citep{croom05}.  
Thus the higher-redshift evolution of the 
luminosity density is entirely in accord with the expectation 
that it is dominated by accretion onto black holes contained
within massive galaxy halos.

At $z<0.5$ the luminosity density falls off faster than predicted,
although even at $z=0$ the predicted and observed values agree to a
factor two.
The implication is that although halos and black holes grow coevally
at higher redshifts, in the more nearby universe black hole growth may
have decoupled from halo growth.
There is growing
evidence that accretion rate in the low-redshift universe 
depends on host galaxy mass, with
lower mass galaxies having black holes that at low redshift are preferentially
growing with respect to their higher-mass counterparts
(e.g. \citealt{heckman}) - so-called
``cosmic downsizing''.  
In detail, to explain cosmic downsizing we may
need to recognise that in the late universe higher mass galaxies have slowed
down their growth more than lower-mass galaxies, perhaps as a result of the
environment they find themselves in at late epochs, or perhaps as a result
of feedback.
An alternative explanation is that the mean radiative
efficiency may fall if AGN accreting at lower Eddington ratios switch
to an ADAF mode of accretion (e.g. \citealt{beckert}).

What we can learn from the PCE model however is that
these processes are secondary to the primary cause of AGN evolution:
the cosmic build-up of structure and its rate of change.
We reiterate that the success of the PCE hypothesis does not mean that
complex physics is not operating: it simply means that averaged across all
massive galaxies, there is a mean net effect which corresponds to black
holes growing at about the same rate as their host dark halos.  Individually
we expect black holes to go through periods of inactivity punctuated by
bursts of accretion probably associated with merger events between galaxies,
remembering that the galaxy mass function itself is also built up
hierarchically by mergers between less massive progenitors.
But just as the average effect on the galaxy mass function is a steady build-up
of mass, so there is a steady build-up of mass in the black hole population.

A final consistency check is to ask whether the inferred local black hole mass 
density could indeed have been built up during the process of luminous
accretion.  This question has largely been answered by \citep{marconi},
who have shown that the \citet{ueda} luminosity function correctly
predicts not only the hard extragalactic X-ray background spectrum but 
also the local black hole mass density, as already discussed above.  The
implication is that not only are the estimates of bolometric luminosity
density consistent with the relic black hole population, but that there
cannot have been any substantially larger amount of ``hidden'' (Compton-thick
or radiatively-inefficient) black hole growth, unless the observed 
AGN have an uncomfortably high radiative efficiency ($\epsilon \gg 0.1$).

\subsection{The evolution in the Eddington ratio of supermassive black holes} 
An alternative view of the evolution expected from the PCE hypothesis is
to calculate the mean Eddington ratio $\langle\lambda\rangle$, where 
$\lambda$ is the ratio of the actual mass accretion rate $\dot{M}_{\rm acc}$
to the Eddington mass accretion rate $\dot{M}_{\rm Edd}$ defined by 
$L_{\rm Edd} = \epsilon \dot{M}_{\rm Edd} c^2$, so that
\begin{eqnarray}
\nonumber
\left\langle\lambda\right\rangle &=& 
\left\langle\frac{\epsilon c \sigma_{\rm T}\dot{M}_{\rm acc}}{4\pi GM_{\rm BH}m_{\rm P}}\right\rangle \\
&=& 
\frac{c \sigma_{\rm T}}{4\pi Gm_{\rm P}}
\left\langle\frac{\epsilon f(M_{\rm H})}{\left(1-\epsilon\right)}
\right\rangle
\left | \frac{d\delta_{c}}{dt} \right |.
\end{eqnarray}
In deriving this expression, it is important to remember that we have
calculated the PCE-predicted average Eddington rate of {\em all} 
black holes in galaxies, not simply those that happen to be visible as
AGN at any particular epoch.  As we don't expect accretion to be a smooth
continuous process for any one galaxy, this is an important distinction.
An individual galaxy with an observed high rate of mass accretion could have
an individual Eddington ratio approaching unity as seen in even local
AGN (e.g. \citealt{onken})
but the whole population of galaxies would have a much lower mean Eddington ratio.

\begin{figure}
  \resizebox{8cm}{!}{ 
  \rotatebox{270}{
  \includegraphics{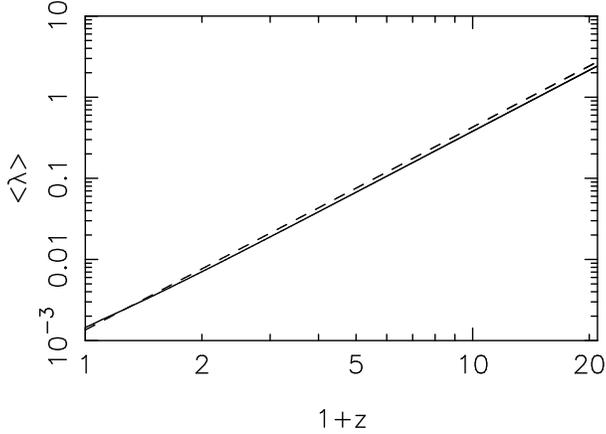}
  }}
  \caption{
Ratio of mean accretion rate to the limiting Eddington rate,
$\lambda$, as a function of $(1+z)$ over the range $0<z<20$,
plotted logarithmically to demonstrate the near power-law dependence.
Results are shown for $M_{\rm H}= 10^{12.5} h^{-1} \msun$ 
and mean radiative efficiency $\langle\epsilon\rangle=0.04$ for:
(a) $\Lambda$CDM cosmology, $\Omega_0=0.3$, $\Omega_{\Lambda}=0.7$, 
$H_0 = 73$\,km\,s$^{-1}$\,Mpc$^{-1}$, $\sigma_8=0.74$ (solid curve);
(b) Einstein-de Sitter cosmology with parameters as in (a) except
$\Omega_0=1$, $\Omega_{\Lambda}=0$ (dashed curve).
}
\label{fig:ER}
\end{figure}

So what significance can we attach to the mean Eddington ratio?
If $\langle\lambda\rangle$ has a high value, it implies
that, on average, there is plenty of matter available to fuel luminous
accreting black holes at or close to their Eddington rate.
In fact, values significantly higher than unity would imply that there
is actually too much accreting matter available: we should expect
that black hole growth by accretion would be limited to the Salpeter rate
but that overall growth by mergers might be important.
Conversely, a low value implies that, on average, there is
insufficient new material accreting onto massive galaxies to maintain
Eddington-limited black hole growth.
Fig.\,\ref{fig:ER} shows $\left\langle\lambda\right\rangle$ 
as a function of redshift for a flat, $\Lambda$CDM cosmology,
assuming $\Omega_M=0.3$, $H_0 = 73$\,km\,s$^{-1}$\,Mpc$^{-1}$, $\sigma_8 = 0.74$,
$M_{\rm H} = 10^{12.5} h^{-1} \msun$, radiative efficiency $\epsilon=0.04$
and without non-linear modification (section\,\ref{sec:pce}).

The result plotted is remarkable in two respects.  As already
discussed, $\langle\lambda\rangle$ depends only weakly 
on the mass of dark matter
halo and on the shape of the dark matter power spectrum, 
with no dependence on the
fraction of a halo's total mass (dark matter plus baryons), that ends
up in the black hole.  Even the dependence on cosmology is weak, with
a near power-law dependence on $1+z$ with exponent $n \simeq 2.4$ 
at $z<1$ for the $\Lambda$CDM
cosmology compared with $n = 2.5$ for Einstein-de Sitter.
Second, in principle there was no reason why the growth of dark matter
halos should have any interesting relationship to the Eddington limit,
except that both are governed by gravity.  Yet it turns out
that at low redshifts the mean ratio 
$\langle\lambda\rangle$ is consistent with the range of
observationally determined
values $\lambda_L$ (determined as the ratio of bolometric luminosity to
estimated Eddington luminosity)
for galaxy samples with a mean value 
$\langle\lambda_L\rangle \sim 0.0035$
\citep{ho04} as discussed below.  
At moderate redshifts the mean ratio increases to a value 
$\langle\lambda\rangle \sim 0.1$ at $z \sim 4$, implying that
at that epoch every galaxy has an actively accreting nuclear
black hole (although typically with lower mass than today's black holes).

\section{Discussion}

\subsection{Dispersion in Eddington ratio}

The above analysis has only enabled us to calculate the mean evolution
in luminosity density and 
in $\langle\lambda\rangle$.  Since accretion events are discrete, we expect
there to be a distribution of $\lambda$ values at any given cosmic
epoch.  In principle one could either attempt to calculate that
distribution from EPS theory (e.g. \citealt{cole}) or
from numerical simulation (e.g. \citealt{hopkinsb}, \citealt{dimatteo}),
or one could look at the
observational constraints on the distribution.  \citet{ho04} has
estimated the distribution of $\lambda_L$ for Palomar galaxies, with
bolometric luminosities estimated from $H\alpha$ and black hole mass
estimated from applying the \citet{tremaine} relation to measured
stellar velocity dispersions.  The distribution at essentially
zero redshift is broad, with a mean $\langle\lambda_L\rangle \sim 0.0035$, 
and although there
are large uncertainties associated with the determination of $\lambda_L$
the agreement to a factor about 2 with the value 
$\langle\lambda\rangle \sim 0.0015$ obtained at $z=0$ for 
$\langle\epsilon\rangle = 0.04$ 
(shown in Fig.\,\ref{fig:ER}) is encouraging.  The
\citet{ho04} values have a large dispersion, which results in there
being a small number of galaxies with $\lambda_L \sim 0.1$ that are
visible as AGN \citep{onken}, and the vast majority of galaxies
accreting at rates closer to $\langle\lambda_L\rangle$.  

We again emphasise that the mean value $\langle\lambda\rangle$
calculated here is the mean for all galaxies and is not the mean
value expected for samples of AGN.  By definition AGN are selected because 
they are actively accreting and they must therefore be biased to higher
values than the mean.  Using the results from this
paper to predict the distribution and evolution of AGN requires
knowledge of or assumptions about the distribution of $\lambda$ 
and will be discussed
further by Babi\'{c} et al. (in preparation). 

\subsection{AGN evolution and accretion at high redshift}

As seen in section\,\ref{sec:lumdensity}, the absolute value and the evolution
of either
the observed space or luminosity density appears to be 
a combination of two competing but inter-related effects:
the build-up of massive halos and hence black holes with decreasing
redshift, and the decreasing accretion rate with decreasing redshift
calculated in this paper.  \citet{croom05} have shown that luminous
QSOs inhabit dark matter halos of mass $\sim 10^{12.5} h^{-1} \msun$:
at low redshifts ($z<1$) the halo mass function, and hence the black hole mass
function, is changing little at this mass, and AGN evolution is
dominated by the evolution in $\lambda$.  In this case we expect the
evolution of the population to be dominated by apparent {\em luminosity
evolution}: on average black holes at lower redshift accrete at a lower
rate than black holes at higher redshift.  Broadly speaking, this
is what is observed \citep{boyle00}, although it's not the whole
story \citep{steffen,cowie,ueda,zheng,barger}. At higher redshifts ($z>2$)
the growth of the black hole mass function is significant and 
probably dominates the evolution, causing a decline in space 
density with redshift (Fig.\,\ref{fig:fig5}).

At $z \sim 4$
$\langle\lambda\rangle$ approaches the values found for the 
most luminous QSOs at
lower $z$: at this epoch in cosmic history the average dark matter
halo is accreting at a rate sufficient to supply close to the
Eddington luminosity of its nuclear black hole: the majority of
galaxies with such a black hole would be active.  At higher redshifts
still the rate of accretion of matter onto dark halos exceeds that
required to supply the Eddington rate.  In this case it is unlikely
that super-Eddington accretion onto black holes would occur: it is
more likely that nuclear outflows would limit the accretion process 
of individual black holes to
about the Eddington rate and it is also possible that the
black-hole/bulge relation may be built up during this phase of galaxy
and black hole growth \citep{king,kingpounds}.

\subsection{Comparison with previous work}
In previous numerical (e.g. \citealt{kauffmann, volonteri,croton}) and
analytic (e.g. \citealt{percival99}) attempts to
understand AGN evolution and supermassive black hole growth, AGN
activity was hypothesised to be directly associated with merger
events, and prescriptions were adopted that describe the relationship
between the halo mass of a galaxy and the resulting AGN's luminosity,
and between the rate of mergers and the evolving space density of AGN.
Implicitly, a relationship between black hole mass and halo mass is assumed.
To obtain the correct space density of AGN it was also necessary to
assume a timescale for the luminous phase of a black hole following a
merger.  In these models, the evolution in EPS merger rate effectively
produces number density evolution, whereas the dominant component of
observed QSO evolution takes the form of luminosity evolution.  As argued in
the introduction, there must in fact have been evolution in
mean accretion rate if local massive galaxies contain dormant
supermassive black holes.  This work shows how that evolution arises
naturally in $\Lambda$CDM models.  There has been no requirement to
make assumptions about the luminosity/halo relation or the link
with mergers and no assumption required about the timescale of AGN
activity.  The calculation presented here {\em has} assumed coeval
growth of black holes: similar assumptions are made in, for example,
\citet{volonteri}, but this assumption certainly needs further testing
and evaluation.  Overall, the extremely good agreement between the
predicted and observed AGN luminosity density and its evolution, without
any ``model-tuning'', points to a close link between AGN cosmological
evolution and the evolution in accretion onto galaxy halos.  
Conversely,
predictions of recent semi-analytic models that attempt to provide
recipes for the complex physics governing black hole growth 
\citep{croton}
predict a peak in black hole formation rate (and hence presumably emitted
luminosity density arising from accretion) at $z \sim 3$, with a
decline to $z=0$ of a factor $\sim 6$, in 
disagreement with the observed luminosity density presented in 
section\,\ref{sec:lumdensity}.

Numerical simulations have also been used to try to understand
the link between black hole growth and galaxy mergers, and in particular
to understand the role of feedback in forming the well-defined relation
between galaxy velocity dispersion and black hole mass
(e.g. \citealt{robertson, dimatteo, hopkinsa, hopkinsb}).  
These approaches are directly complementary to the
question addressed in this paper, namely to understand analytically the
effect that the cosmic slowdown in halo growth has on the typical
black hole accretion rate.  It may be that combining the two approaches
will finally lead us to a detailed understanding of AGN evolution.

\section{Conclusions}
The results from this work may be summarised as follows.
\begin{enumerate}
\setlength\itemsep{0em}

\item
Press-Schechter theory may be used to calculate analytically the mean
mass accretion rate of dark matter halos.  The result of the calculation
is in good agreement with previous numerical work but offers a significant
improvement over previously-published fits to numerical simulations.
The mass accretion rate depends almost linearly on mass but is very insensitive
to choice of cosmological parameters within the range normally considered.
We present fitting formulae to allow easy calculation of the accretion
rate.

\item
It seems likely that the deduced strong decline in halo mass accretion will
have an effect not only on the build-up of dark halos but also on the
baryonic structures they contain.  There may therefore be a close link
between the cosmological evolution in star formation rate, in AGN
accretion and dark halo accretion.

\item
We have investigated arguably the simplest hypothesis one could
make for black hole growth, ``Pure Coeval Evolution'' (PCE),
which postulates that, {\em on average}, 
supermassive black holes have growth that tracks
the growth of their parent dark halos.  In this case it is straightforward to
calculate the expected mass accretion rate onto black holes and, 
estimating the local black hole mass density, we predict a
value for the integrated AGN luminosity density 
in remarkable agreement at $z \ga 0.5$
with the value deduced from hard X-ray surveys.

\item
The evolution in integrated AGN luminosity
density at $z > 0.5$ is well-matched by this model 
if the integrated black hole mass density tracks the
mass density in massive $M_{\rm H}>10^{11.5}\msun$ halos, 
consistent with
the value obtained by inferring the clustering bias of QSOs in 
the 2dF QSO redshift survey \citep{croom05}

\item
At $z<0.5$ the observed luminosity density falls off faster than
predicted, although even by $z=0$ the overprediction is only a factor
two.  It seems that in the recent universe the black hole growth
has started to decouple from dark halo growth, although a decrease in
mean radiative efficiency associated with an increasing prevalence of ADAFs
may also occur.

\item
Expressed in terms of the Eddington ratio, at sufficiently high redshifts
($z \ga 4$), halos would have black holes accreting at a
significant fraction of the Eddington rate - we expect all massive galaxies
to contain an active AGN.  At higher redshifts still, the nominal Eddington
ratio could greatly exceed unity.  The response of the black hole would be
to limit its growth to the Salpeter rate, but such systems might produce
significant outflows \citep{king,kingpounds} which in turn would produce
a significant feedback effect on their hosts.  It is also likely that in this
phase of black hole growth, growth by mergers between halos is faster than
growth by accretion.  Merger-dominated growth is one way of avoiding the
problem of high-mass black holes at high redshift that have had insufficient
time to grow at the Salpeter rate from solar- or intermediate-mass 
black hole progenitors \citep{willott}.
\end{enumerate}

\noindent
Further work will model the AGN luminosity function and the contribution to
the X-ray background (Babi\'{c} et al. in preparation).  

\acknowledgements{
We are grateful to Y.\,Ueda for supplying data from \citet{ueda}.
AB acknowledges support from the Clarendon Fund.}

\appendix

\section{Calculating $\mathbf{\lowercase{d}\delta_{\lowercase{c}}/\lowercase{dt}}$}
  \label{app:A}

In this section we discuss the calculation of $d\delta_c/dt$ and approximations
to it for cosmologies with a cosmological constant (the analysis could be extended
to include more general dark energy cosmologies following the approach of
Percival (2005) but this is beyond the scope of this appendix).  
We consider quantities that evolve as a
function of scale factor $a=(1+z)^{-1}$. If no dependence is quoted, the quantities
should be assumed to be calculated at $a=1$ ($z=0$).  

The critical overdensity for collapse $\delta_c(a)$ in the
spherical top-hat collapse model is the linear overdensity, extrapolated
to the present day, that leads to the collapse 
of a homogeneous spherical region
to a singularity at scale factor $a$.
This is used
in EPS theory to link overdensities with their predicted collapse
times. However, two alternative formalisms are often
considered for the collapse of perturbations in EPS theory.  
\begin{enumerate}
  \item The overdensity field is assumed to grow with the linear
  growth factor, and when perturbations reach a particular critical
  overdensity they are said to have collapsed. This suggests that,
  given $\delta_c$, we can calculate $\delta_c(a)=\delta_cD/D(a)$,
  where $D(a)$ is the linear growth factor. 
  \item An overdense region is
  considered to be spherical and to evolve according to the Friedmann
  equations, and hence its collapse time may be calculated. The
  overdensity value is extrapolated to the epoch at which the density
  field is normalised (conventionally, the present day) using the linear 
  growth factor.
\end{enumerate}

\begin{figure}
  \resizebox{8cm}{!}{ 
  \rotatebox{0}{
  \includegraphics{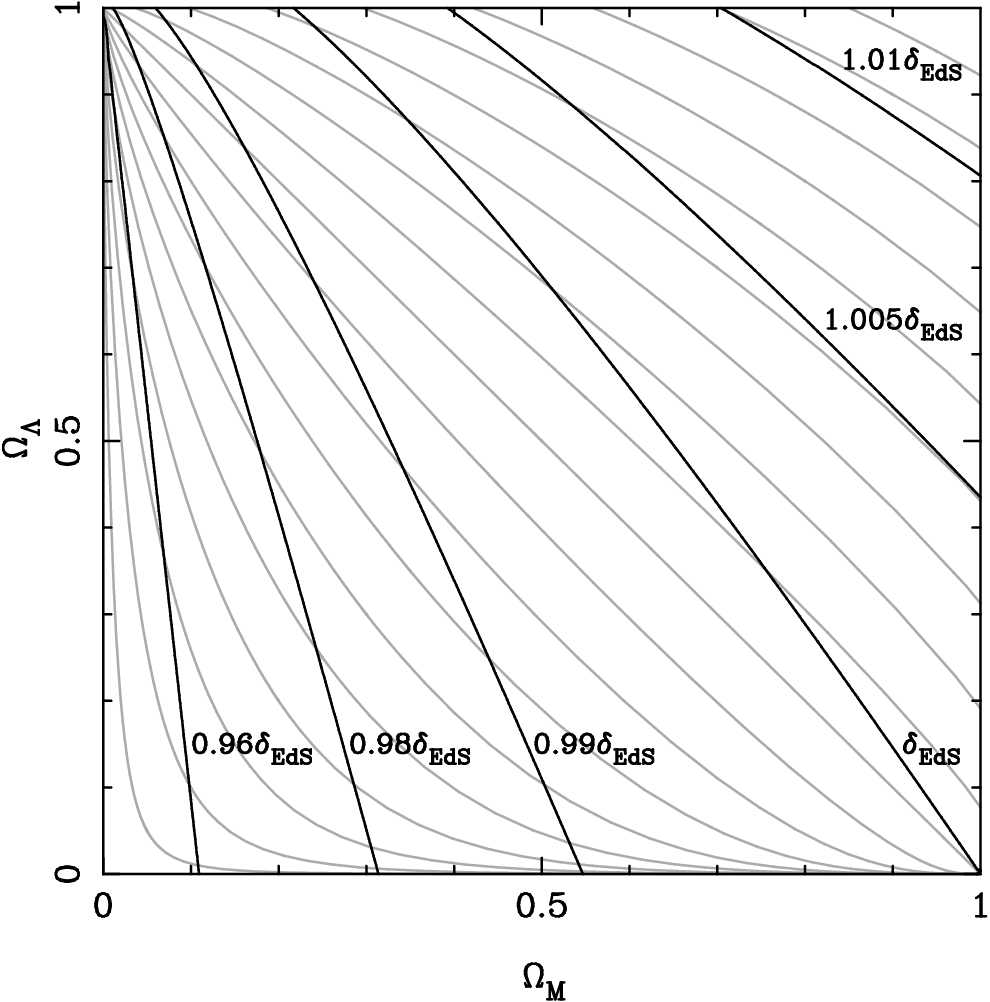}}}
  \caption{Plot showing the mismatch between contours of constant
  $\delta_c$ (black lines) and the lines along which $\Omega_M$ and
  $\Omega_\Lambda$ vary with the evolution of the universe (grey
  lines). The contours of constant $\delta_c$ are plotted as a
  ratio of $\delta_{\rm EdS}$, the critical overdensity for collapse at
  redshift zero in an Einstein-de Sitter cosmology. Because these
  lines cross, the evolution of $\delta_c(a)$ is not solely dependent
  on the linear growth factor. \label{fig:appfig1}}
\end{figure}

These two alternatives give different formulae for
$\delta_c(a)$. In Fig.~\ref{fig:appfig1} we plot
contours of constant $\delta_c$ (black lines) within the plane of
($\Omega_M,\Omega_\Lambda$) values, calculated using the method of
\citet{percival00}. Contours are plotted as a function of the value of 
$\delta_c$ in an Einstein-de Sitter cosmology:
\be
  \delta_{\rm EdS} = \frac{3}{20}(12\pi)^{2/3} \simeq 1.686 \,.
\ee
We also plot lines along which universes evolve (grey lines) given by 
\begin{eqnarray}
  \Omega_M(a) & = & 
    \frac{\Omega_M}{a+(1-a)\Omega_M+(a^3-a)\Omega_\Lambda}\\
  \Omega_\Lambda(a) & = & 
    \frac{a^3\Omega_\Lambda}{a+(1-a)\Omega_M+(a^3-a)\Omega_\Lambda}.	
\end{eqnarray}	
Because $\delta_c$ changes along the lines of evolution, 
we see that the
spherical top-hat collapse model does not predict evolution of the
critical overdensity given by $\delta_c(a)=\delta_cD/D(a)$. 

The first picture of EPS theory as corresponding to a growing field
does not follow from the spherical top-hat collapse model. However,
Fig.~\ref{fig:appfig1} shows that the evolution of $\delta_c$ along
lines of evolving ($\Omega_M(a),\Omega_\Lambda(a)$) is small. In the
remainder of this appendix we consider the error introduced in
$|d\delta_c(a)/dt|$ by assuming that the evolution in $\delta_c(a)$ is
governed solely by the linear growth factor. The linear growth factor
$D(a)$ is given by
\be
  D(a) = \frac{5}{2}\Omega_M H_0^2 H(a) \int^a_0\frac{da}{\left(aH\left(a\right)\right)^3},
    \label{eq:Dexact}
\ee
where
\be
  \left(\frac{1}{a}\frac{da}{dt}\right)^2 = H(a)^2=
    H_0^2(\Omega_\Lambda+\Omega_Ma^{-3}-(\Omega_m+\Omega_\Lambda-1)a^{-2}).
    \label{eq:H}
\ee
$D(a)$ may be approximated using the fitting formula of
Carroll et al. (1992):
\begin{eqnarray}
  \nonumber
  \lefteqn{D(a) \simeq \frac{5\Omega_M(a)a}{2}
  \left[\Omega_M(a)^{4/7}-\Omega_\Lambda(a)\phantom{\frac{1}{1}}\right.}\\
  & & \hspace{1.0cm}\left. +\left(1+\frac{\Omega_M(a)}{2}\right)
    \left(1+\frac{\Omega_\Lambda(a)}{70}\right) \right]^{-1}.
    \label{eq:Dapprox}
\end{eqnarray}
Eq.~\ref{eq:Dexact} can be differentiated to obtain
\begin{eqnarray}
  \nonumber
  \lefteqn{\frac{dD(a)}{da}=\frac{H_0^2}{H(a)^2}
    \left[\frac{5}{2}\frac{\Omega_M}{a^3}
    -\frac{3}{2}\frac{D(a)\Omega_M}{a^4}\right.}\\
  & & \hspace{1.5cm} \left.+(\Omega_M+\Omega_\Lambda-1)\frac{D(a)}{a^3}
    \right].
    \label{eq:dDda}
\end{eqnarray}
Finally, the approximation to $d\delta_c(a)/dt$ is given by
\be
  \frac{d\delta_c(a)}{dt} \simeq \delta_{\rm EdS} \frac{D}{D(a)^2}
    \frac{dD(a)}{da}\frac{da}{dt}.
  \label{eq:dddt}
\ee

\begin{figure}
  \resizebox{8cm}{!}{ 
  \rotatebox{0}{
  \includegraphics{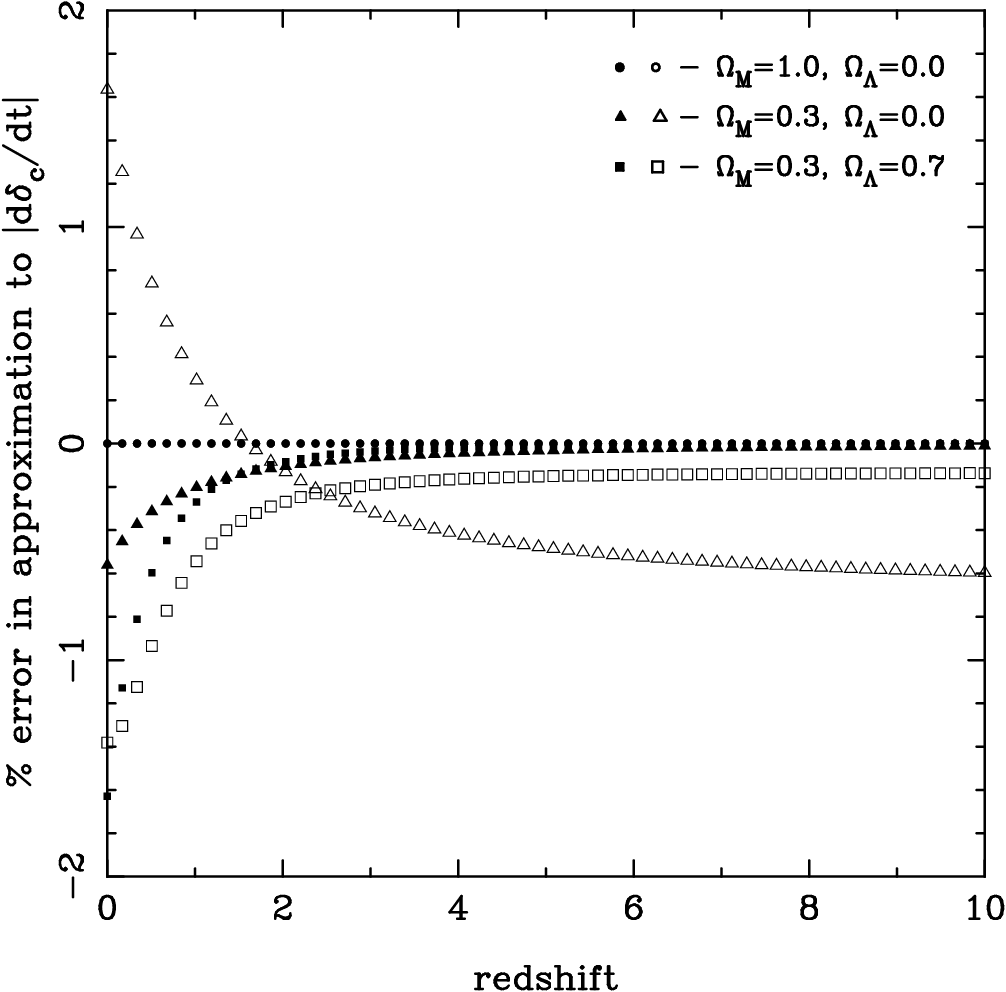}}}
  \caption{The percentage error in approximating the time
  derivative of $\delta_c(a)$ as $\delta_c(a)=\delta_{\rm EdS}D/D(a)$,
  where $D(a)$ is the linear growth factor. Open symbols incorporate
  the approximation for $D(a)$ given by Eq.~\ref{eq:Dapprox}, while
  solid symbols calculate $D(a)$ by numerically integrating
  Eq.~\ref{eq:Dexact}. Results are shown for three cosmological
  models: for an Einstein-de Sitter cosmology, both approximations
  reduce to the exact behaviour. \label{fig:appfig2}}
\end{figure}

Equations~\ref{eq:H}, \ref{eq:Dapprox}, \ref{eq:dDda} \&~\ref{eq:dddt}
combine to provide an analytic approximation to $d\delta_c/dt$. In
Fig.~\ref{fig:appfig2}, we plot the percentage error introduced by this
approximation compared with numerically integrating the 
behaviour of $\delta_c(a)$, calculated using the method of
\citet{percival00} (open symbols). We also consider calculating
$D(a)$ exactly and use this instead of the Carroll et al. (1992)
approximation in Eq.~\ref{eq:dDda} (solid symbols). The solid
symbols show the error in $d\delta_c/dt$ from ignoring the
evolution in $\delta_c$ because of changing $\Omega_M$ and
$\Omega_\Lambda$. Comparison of solid and open symbols shows the error
introduced by Eq.~\ref{eq:Dapprox}. Errors are plotted for three
cosmologies: for the Einstein-de Sitter cosmology, the approximations
reduce to the exact result. For the $\Lambda$ and
open cosmologies, the maximum error is less than 2\%.

\label{lastpage}

\end{document}